\documentclass[12pt,twoside]{article}
\usepackage{amssymb}
\usepackage{amsmath}
\begin{document}
\begin{center}
\LARGE Communication cost of simulating quantum correlations in a class of integer spin singlet states.
\end{center}
\begin{center}
{\bf Ali Ahanj\footnote{Electronic address: ahanj@physics.unipune.ernet.in} and Pramod Joag\footnote{Electronic address : pramod@physics.unipune.ernet.in}} \\
\end{center}
\begin{center}
\textbf{Department of Physics, University of Pune,  Pune - 411007, India}
\end{center}
\begin{flushright}
\end{flushright}
\small {We give a classical protocol to exactly simulate quantum correlations implied by a spin $s$ singlet state for the infinite sequence of \textit{integer} spin $s={1,4,13,...}$ satisfying $(2s+1)=3^{n}$,where $n$ is a positive integer.The required amount of communication is found to increase as $log_{3}d^{2}$ where $d=2s+1$ is the dimension of the spin $s$ Hilbert space.}\\


It is well known that quantum correlations implied by an entangled quantum state of a bipartite quantum system cannot be produced classically, i.e.using only the local and realistic properties of the subsystems, without any communication between the two subsystems[1].By quantum correlations we mean the statistical correlations between the outputs of measurements independently carried out on each of the two entangled parts.Naturally the question arises as to the minimum amount of classical communication (number of cbits) necessary to simulate the quantum correlations of an entangled bipartite system.This amount of communication quantifies the nonlocality of the entangled bipartite quantum system.It also helps us gauge [2] the amount of information hidden in the entangled quantum system itself,in some sense, the amount of information that must be space-like transmitted, in a local hidden variable model,in order for nature to account for the excess quantum correlations.\\
In this scenario, Alice and Bob try and output $\alpha$  and $\beta$ respectively,through a classical protocol,with the same probability distribution as if they shared the bipartite entangled system and each measured his or her part of the system according to a given random Von Neumann measurement.As we have mentioned above, such a protocol must involve communication between Alice and Bob, who generrally share finite or infinite number of random variables. The amount of communication is quantified[3] either as the average number of cbits
$\overline {C}(P)$ over the directions along which the spin components are measured (average or expected communication) and worst case communication, which is the maximum amount of communication $C_{w}(P)$  exchanged between Alice and Bob in any particular execution of the protocol. The third method is asymptotic communication i.e. the limit $lim_{n\rightarrow\infty}\overline{C}(P^{n})$ where $P^{n}$ is the probability distribution obtained when $n$ runs of the protocal carried out in parallel i.e. when the parties recieve $n$ inputs and produce $n$ outputs in one go. Note that,naively,Alice can just tell Bob the direction of her measurement to get an exact classical simulation, but this corresponds to  an infinite amount of communication. the question whether a simulation can be done with finite amount of communication was raised independently by Maudlin[4],Brassard,Cleve and Tapp[5],and Steiner[6].Brassard,Cleve and Tapp used the worst case communication cost while Steiner used the average. Steiner's model is weaker as the amount af communication in the worst case can be unbounded although such cases occur with zero probability.Brassard ,Cleve and Tapp give a protocol to simulate entanglement in a singlet state  EPR pair using six cbits of communication.Toner and Bacon[7] give a protocol to simulate two qubit singlet state entanglement using only one cbit of communication.Interestingly, quantum correlations that cannot be classically simulated without communication also occur in a scenario where incompatible observables are successively measured on class of input (single particle) spin-s states which can be simulated with  a classical protocol with communication between successive measurements[8].\\

Untill now,an exact classical simulation of quantum correlations is accomplished only for spin $1/2$ singlet state,requiring 1 cbit of classical communication[7].It is important to know how does this communication cost increase with $s$,in order to quantify the advantage offered by quantum communication over the classical one.Further,this communication cost quantifies,in terms of classical resourses,the variation of the nonlocal character of quantum orrelations with spin value.An enhancement in this direction was made in [9] where an exact classical protocol to classical simulate spin $s$ singlet state correlations for infinite sequance of spins satisfying $2s+1=2^{n}$,($n$ positive integer) was given.\\
In this paper we give a classical protocol to simulate entanglement in a singlet spin-s state of a bipartite spin s system satisfying $2s+1=3^{n}$,which is an infinite sequence of integer spin values$s={1,4,13,...}$, using $2n$ cbits of communication and $3n$ shared random variables, where $n$is a positive integer.The quantum correlations  
$ \left\langle \alpha \beta\right\rangle $ for a singlet state are given by
\begin{eqnarray}
\left\langle \alpha \beta\right\rangle=-\frac{1}{3} s(s+1)\hat{a}.\hat{b}
\end{eqnarray}
 where $\hat{a}$ and $\hat{b}$ are the unit vectors specifying the directions along which the spin components are measured by Alice and Bob respectively[10].Note that, by virtue of being a singlet state,$ \left\langle \alpha \right\rangle=0=\left\langle\beta\right\rangle$ irrespective of directions $\hat{a} $ and $\hat{b}$.The protocol proceeds as follows:\\ Alice outputs
\begin{eqnarray}
 \alpha=-\sum^{n}_{k=1}3^{n-k}\Theta_{1}(\hat{a}.\hat{\lambda}_{k})
 \end{eqnarray}
 Where:
\begin{eqnarray}
\Theta_{1}(x)= \left\{
\begin{array}{c c }
1  & 1/3 \leq x \leq 1 \\
0  & -1/3 < x  < 1/3 \\
-1 & -1  \leq x  \leq -1/3\\
\end{array} \right.
\end{eqnarray}
so that $\alpha$ can take $3^{n}=2s+1$ values,between $(3^{n}-1)/2 $ and $ -(3^{n}-1)/2 $(we take $\hbar=1$). 
 Alice sends $2n$ cbits $c_{1},c_{2},..,c_{2n}$ to Bob where    $c_{2k-1}=sgn(\hat{a}.\hat{\lambda}_{k}) sgn(\hat{a}.\hat{\mu}_{2k-1})$ and
 $c_{2k}=sgn(\hat{a}.\hat{\lambda_{k}})sgn(\hat{a}.\hat{\mu}_{2k})$; $k=1,2,...,n$.
  $\{\hat{\lambda_{k}}\}$,$\{\hat{\mu_{j}}\}$ ;$k=1,2,...,n$ $j=1,2,...,2n$
are the 3n shared random variables between Alice and Bob.We have used the sgn function defined by $sgn(x)=+1~~ $if~$ x\geq0 $ and $sgn(x)=-1~~ $if~$ x<0$. After receiving these $2n$ cbits from Alice, Bob outputs ~
\begin{eqnarray} \beta&=&\sum^{n}_{k=1}3^{n-k}\Theta_{2}[\hat{b}.(c_{2k-1}\hat{\mu}_{2k-1}+c_{2k}\hat{\mu}_{2k})]\nonumber\\
&=&\sum^{n}_{k=1}3^{n-k}\sum^{1}_{d_{2k-1}=-1}\frac{1+c_{2k-1}d_{2k-1}}{2}\sum^{1}_{d_{2k}=-1}\frac{1+c_{2k}d_{2k}}{2}\Theta_{2}[\hat{b}.(d_{2k-1}\hat{\mu}_{2k-1}+d_{2k}\hat{\mu}_{2k})]
\end{eqnarray}
These sums over $d_{2k-1}$ and $d_{2k}$  have sixteen terms eight out of which are zero.This gives:
\begin{eqnarray}
\beta&=&\frac{1}{2}\sum^{n}_{k=1}3^{n-k}\{c_{2k-1}[\Theta_{2}(\hat{b}.(\hat{\mu}_{2k-1}+\hat{\mu}_{2k}))+\Theta_{2}(\hat{b}.(\hat{\mu}_{2k-1}-\hat{\mu}_{2k}))\nonumber\\
&&-\Theta_{2}(\hat{b}.(-\hat{\mu}_{2k-1}+\hat{\mu}_{2k}))-\Theta_{2}(\hat{b}.(-\hat{\mu}_{2k-1}-\hat{\mu}_{2k}))]+c_{2k}[\Theta_{2}(\hat{b}.(\hat{\mu}_{2k-1}+\hat{\mu}_{2k}))\nonumber\\
&&-\Theta_{2}(\hat{b}.(\hat{\mu}_{2k-1}-\hat{\mu}_{2k}))+\Theta_{2}(\hat{b}.(-\hat{\mu}_{2k-1}+\hat{\mu}_{2k}))-\Theta_{2}(-\hat{b}.(\hat{\mu}_{2k-1}-\hat{\mu}_{2k}))]\}
\end{eqnarray}
In each bracket two terms have same value  as the remaining ones,therefore
\begin{eqnarray}
\beta&=&\frac{1}{2}\sum^{n}_{k=1}3^{n-k}\{c_{2k-1}[\Theta_{2}(\hat{b}.(\hat{\mu}_{2k-1}+\hat{\mu}_{2k}))+\Theta_{2}(\hat{b}.(\hat{\mu}_{2k-1}-\hat{\mu}_{2k}))]\nonumber\\ 
&&+c_{2k}[\Theta_{2}(\hat{b}.(\hat{\mu}_{2k-1}+\hat{\mu}_{2k}))-\Theta_{2}(\hat{b}.(\hat{\mu}_{2k-1}-\hat{\mu}_{2k}))]\}
\end{eqnarray}
Where
\begin{eqnarray}
\Theta_{2}(x)= \left\{
\begin{array}{c c }
1  & 2/3 \leq x \leq 2 \\
0  & -2/3 < x  < 2/3 \\
-1 & -2  \leq x  \leq -2/3\\
\end{array} \right.
\end{eqnarray}
We now prove the protocol by showing that it produces correct expectation values.The expectation value is defined by:
\begin{eqnarray}
E(x)=\frac{1}{(4\pi)^{3n}}\int d\lambda_{1}...d\lambda_{n}d\mu_{1}...d\mu_{2n}x
\end{eqnarray}
Each party's output changes $\Theta$ under the symmetry $\hat{\lambda_{k}}\leftrightarrow -\hat{\lambda_{k}}$ and $\hat{\mu_{j}}\leftrightarrow -\hat{\mu_{j}}$
;$k=1,2,...,n$ and $j=1,2,...,2n$ so $ \left\langle \alpha \right\rangle=0=\left\langle\beta\right\rangle$ as all $\hat{\lambda}$ s and $\hat{\mu}$ s are uniformly distributed.The joint expectation value 
$\left\langle \alpha\beta \right\rangle$ can be calculated as follows. We have \\
\begin{eqnarray}
\left\langle \alpha\beta \right\rangle =-\sum^{n}_{k,l=1}3^{2n-k-l}E\{\Theta_{1}(\hat{a}.\hat{\lambda}_{k}) 
\Theta_{2}[\hat{b}.(c_{2l-1}\hat{\mu}_{2l-1}+c_{2l}\hat{\mu}_{2l})]\}
\end{eqnarray}
As shown in [9],only those terms are nonzero that satisfy $k=l$ 
,therefore only the terms with $k=l$ survive in $\left\langle \alpha\beta \right\rangle$ so that
\begin{eqnarray}
\left\langle \alpha\beta \right\rangle =-\sum^{n}_{k=1}3^{2n-2k}E\{\Theta_{1}(\hat{a}.\hat{\lambda}_{k})\Theta_{2}[\hat{b}.(c_{2k-1}\hat{\mu}_{2k-1}+c_{2k}\hat{\mu}_{k})]\}
\end{eqnarray}
By substituting (5) in the above equation and using the fact that the resulting four terms have the same expectation as they are related by the symmetries $\hat{\mu_{2k-1}}\leftrightarrow \hat{\mu_{2k}}$ or $\hat{\mu_{2k-1}}\leftrightarrow -\hat{\mu_{2k}}$ we find:
\begin{eqnarray} 
 \left\langle \alpha\beta \right\rangle =-\sum^{n}_{k=1}3^{2n-2k}E\{2\Theta_{1}(\hat{a}.\hat{\lambda}_{k})\Theta_{2}[\hat{b}.(\hat{\mu}_{2k-1}+\hat{\mu}_{2k})]\}
\end{eqnarray}
 The expectation value in eq(11) is evaluated in Appandix. The result is 
 \begin{eqnarray}
  \left\langle \alpha\beta \right\rangle=-\frac{2}{3}\hat{a}.\hat{b}\sum^{n}_{k=1}3^{2n-2k}
  \end{eqnarray}
 Summing the geometric series and using $(2s+1)=3^{n}$ we get
 \begin{eqnarray}
  \left\langle \alpha\beta \right\rangle=-\frac{1}{3}s(s+1)\hat{a}.\hat{b}
  \end{eqnarray}
 The above protocol applies to infinite, although sparse, subset of the set of all spins (all integral and half integral values). The most important finding is that the amount of worst casecommunication goes as $log_{3}(2s+1)^2$ or as $log_{3}s^2$ for $s>>1$.Since the nonlocal correlations get stronger with the Hilbert space dimension $d=2s+1$ [3],we expect that the amount of communication required to classically simulate the quantum correlations to increase with $d$.Our result that the required communication diverges logarithmically is consistant with such an expectation. Further, this logarithmic increase in the required communication for the classical simulation of the singlet state correlations with $d$,enables us to conjecture that the required communication would monotonically increase with $d$, for all spins.At any rate,this result points towards a quantitative relationship between the quantum and classical communication complexities.\\
 \newpage
 \textbf{Appendix}
 
 The average value in equation(11)is evaluted as follows
  \begin{eqnarray}
 &&E\{2\Theta_{1}(\hat{a}.\hat{\lambda}_{k})\Theta_{2}[\hat{b}.(\hat{\mu}_{2k-1}+\hat{\mu}_{2k})]\}=
 \frac{2}{(4\pi)_{3}}\int d\hat{\lambda}_{k}\Theta_{1}(\hat{a}.\hat{\lambda}_{k})sgn(\hat{a}.\hat{\lambda}_{k})\nonumber\\
&& \int d\mu_{2k-1}sgn(\hat{a}.\hat{\mu}_{2k-1})\int d \mu_{2k}\Theta_{2}(\hat{b}.(\hat{\mu}_{2k-1}+\hat{\mu}_{2k})) 
\end{eqnarray}
We first integrate over $\hat{\mu}_{2k}$,taking $\hat{b}$ to point along the positive z axis.
\begin{eqnarray}
&&\int d \mu_{2k}\Theta_{2}(\hat{b}.(\hat{\mu}_{2k-1}+\hat{\mu}_{2k}))=\int sin\theta_{2k}d\theta_{2k}d\phi_{2k}(cos\theta_{2k-1}+cos\theta_{2k})=\nonumber\\
&&(-2\pi)[-(\frac{1}{3}+cos\theta_{2k-1})+1]+(2\pi)[-(\frac{1}{3}-cos\theta_{2k-1})+1]\nonumber\\
&&=4\pi cos\theta_{2k-1}=4\pi(\hat{b}.\hat{\mu}_{2k-1})
\end{eqnarray}
We now take $\hat{a}$ to point along the positive $z$ axis, set $\hat{b}=(sint,0,cost)$, and integrate over $\hat{\mu}_{2k-1}$,obtaining
\begin{eqnarray}
&&\int d\hat{\mu}_{2k-1}sgn(\hat{a}.\hat{\mu}_{2k-1})\hat{b}.\hat{\mu}_{2k-1}=(2\pi)cost\int sgn(cos\theta_{2k-1})cos\theta_{2k-1}sin\theta_{2k-1}d\theta_{2k-1}\nonumber\\
&&=(2\pi)cost=(2\pi)\hat{a}.\hat{b}
\end{eqnarray}
The last integral is evaluated seperable:
\begin{eqnarray}
\int d\lambda_{k}\Theta_{1}(\hat{a}.\hat{\lambda}_{k})sgn(\hat{a}.\hat{\lambda}_{k})=(4\pi)(\frac{2}{3})
\end{eqnarray}
so,the expectation value is obtained by (15),(16)and(17)
\begin{eqnarray}
E\{2\Theta_{1}(\hat{a}.\hat{\lambda}_{k})\Theta_{2}[\hat{b}.(\hat{\mu}_{2k-1}+\hat{\mu}_{2k})]\}=(\frac{2}{3})\hat{a}\hat{b}
\end{eqnarray}

\newpage
\textbf{Refrence}

 \begin{verse} 
 [1]~ J.S.Bell Physics (Loug Islaud City,N.Y)\textbf{1},195(1964).

 [2]~ A.A.M\'ethot, quant-ph/$0304122 $

 [3]~ S.Pironio.Phys.Rev.A. \textbf{68},062102(2003).
 
 [4]~ T.Maudlin,in PSA 1992,Volume 1,edited by D.Hull,M.Forbes,and K.Okruhlik(Philosophy
 
 ~~~~ of science Association,East Lansing,1992),pp,404-417
 
 [5]~ G.Brassard,RR.Cleve,and A.Tapp,Phys.Rev.Lett.\textbf{83},1874(1999).
 
 [6]~ M.Steiner,Phys,Lett.A\textbf{ 270},239(2000).
 
 [7]~ B.F.Toner and D.Bacon,Phys.Rev.Lett.\textbf{91},187904(2003).
 
 [8]~ A.Ahanj and P.Joag,quant-ph/0602005.
 
 [9]~ A.Ahanj and P.Joag,quant-ph/0603053.
 
 [10]~A.Peres,{\it Quantum Theory:Concepts and Methods},(Kluwer Academic Publishers~ 1993).
 \end{verse}

\end{document}